\documentclass[%
 reprint,
 amsmath,amssymb,
 aps,
]{revtex4-1}
\usepackage{multirow}
\usepackage{float}
\usepackage{flushend}

\usepackage{graphicx}% Include figure files
\usepackage{dcolumn}% Align table columns on decimal point
\usepackage{bm}% bold math
\usepackage{multirow}
\begin{document}

%\preprint{APS/123-QED}
\title{Dislocations as a boundary between charge density wave and oxygen rich phases in a cuprate high temperature superconductor}% Force line breaks with \\

\author{Nicola Poccia}
 \affiliation{Department of Physics, Harvard University, Cambridge, Massachusetts 02138, USA}
  \email{npoccia@physics.harvard.edu}
 
\author{Alessandro Ricci}
\affiliation{Deutsches Elektronen-Synchrotron DESY, Notkestraße 85, D-22607 Hamburg}
\email{phd.alessandro.ricci@gmail.com}

\author{Gaetano Campi}
\affiliation{Institute of Crystallography, CNR, via Salaria Km 29.300, Monterotondo, Roma, 00015, Italy}
\email{gaetano.campi@ic.cnr.it}

\author{Antonio Bianconi}
\affiliation{RICMASS Rome International Center for Materials Science Superstripes, via dei Sabelli 119A, 00185 Roma, Italy;\\
   Institute of Crystallography, CNR, via Salaria Km 29.300, Monterotondo, Roma, 00015, Italy;\\
    National Research Nuclear University, MEPhI, Kashirskoe sh. 31, 115409, Moscow, Russia}
    \email{antonio.bianconi@ricmass.eu}

%            %  but any date may be explicitly specified

\begin{abstract}
Multiple functional ionic and electronic orders are observed in high temperature superconducting cuprates. The charge density wave order is one of them and it is spatially localized in spatial regions of the material. It is also known that the oxygen interstitials introduced by chemical intercalation self-organize in different oxygen rich regions corresponding with hole rich regions in the Cu$O_2$ layers left empty by the charge density wave order domains. However, what happens in between these two order is not known, and neither there is a method to control this spatial separation. Here we demonstrate by using scanning nano X-ray diffraction, that dislocations or grain boundaries in the material can act as boundary between charge density wave and oxygen rich phases in a optimally doped La$_2$CuO$_4$$_+$$_y$ high temperature superconductor. Dislocations can be used therefore to control the anti-correlation of the charge density wave order with the oxygen interstitials in specific portion of the material. 

\end{abstract}

\pacs{Valid PACS appear here}% PACS, the Physics and Astronomy
%\keywords{Suggested keywords}%Use show keys class option if keyword
                              %display  desired
\maketitle

%\tableofcontents

Dislocations or grain boundaries are known to change the properties of electronic materials \cite{Setter2000} \cite{Hilgenkamp2002}. Although dislocations are often considered as a problem for the performances of a material, it is from the very beginning of semiconductor science that scientists tried to use them to engineer their functionalities \cite{Calzecchi1968}. Currently, novel methods of fabrication to realize networks of dislocation have been adopted to change the electronic functionalities of Si \cite{Kittler2007}. In the proximity of dislocations, the strain of a material is modulated \cite{Davletova2008}. Strain (e.g. chemical pressure, microstrain, misfit-strain) is a material dependent parameter which have a deep influence in the electronic properties of the material and it has been proposed to play a crucial role also in the properties of high temperature superconductors \cite{Poccia2010} \cite{bia2000}as well for their phase diagram description \cite{Kugel2009}. In cuprates, since the spacer layer  has a dissimilar lattice constants than the active layer, the mismatch typically results in the formation of strain-relieving misfit dislocations \cite{Venables1984}. However, the search for an accurate description of the strain in strongly disordered systems is a current field of research \cite{Egami2011} \cite{Campi2006} \cite{Bozin2016} \cite{Campi2012}.

\begin{figure*}[t!]
	\begin{center} \includegraphics[width=0.9\linewidth] {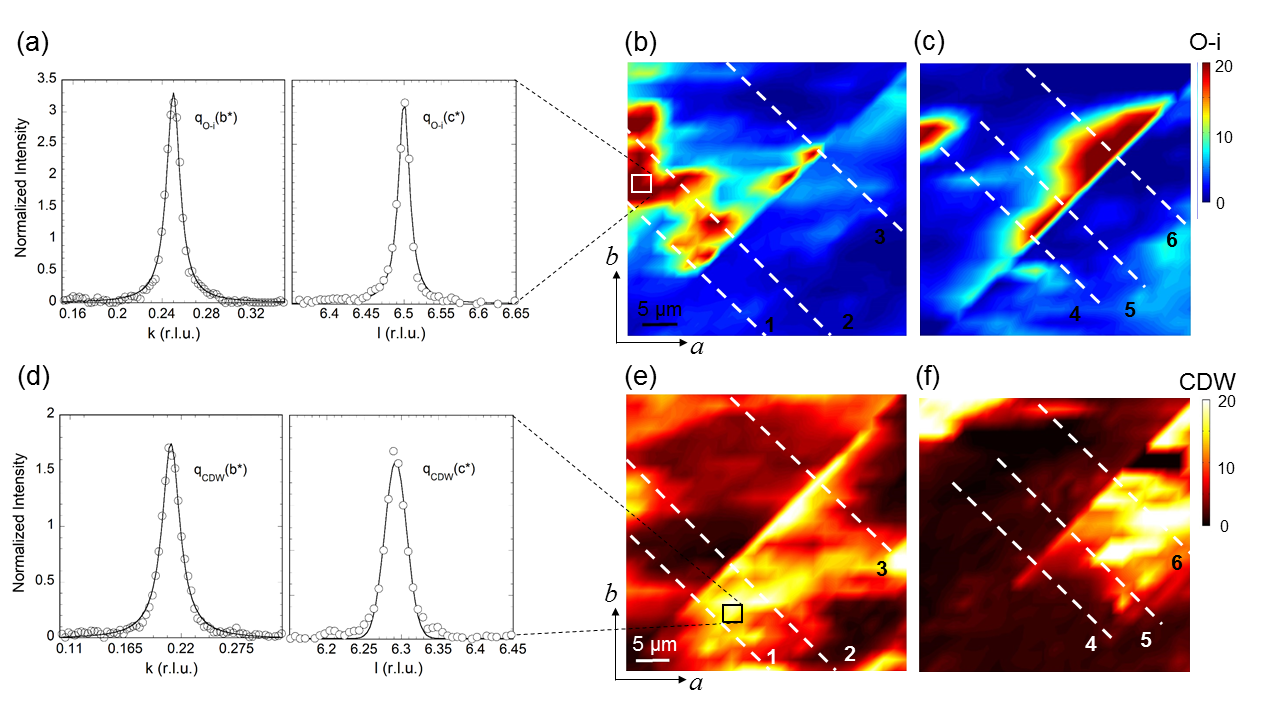}
		\caption{Panel (\textbf{a}) shows the representative X-ray scattering profiles 
		of a typical O-i satellite due to ordered oxygen-interstitials  with wave vector q$_O$$_-$$_i$. 
		The solid lines are the Lorentzian best fit curves.
		The panels (\textbf{b}) and (\textbf{c}) show the xy map of the intensity of the O-i satellite 
		in different portions of the sample. Two different diagonal dislocations which induce sharp boundaries are shown. Here the color goes from dark blue to dark red as indicated in the O-i color bar.
		Panel (\textbf{d}) shows representative scattering 
		profiles of the satellite due to charge density wave order with wave-vector q$_{CDW}$.
		 Panel (\textbf{e}) and (\textbf{f}) show the xy mapping of the intensity of the CDW satellite in the same  spatial
		portions shown in panels  respectively (\textbf{b}) and (\textbf{c}). 
		Here the color goes from dark red to white as indicated in the CDW color bar.
		Visual inspection of the images shows that close to the diagonal linear boundary,  the CDW or O-i order abruptly changes. Comparing panels (\textbf{b}) and  (\textbf{c}) with the panels (\textbf{e}) and  (\textbf{f}), it is clear that the dislocation plays the role of boundary between the O-i rich and CDW rich regions of the sample. The bar is 5 micron. }
		\label{criticality}
	\end{center}
\end{figure*}

Dislocations are a manifestation of a strongly disordered system and 
recently they have been exploited as resistive switching in oxides \cite{Szot2006}, since
 they can act as bistable nanowires. The lattice can be sensitively 
 different at the core of a dislocation \cite{Jia2005}, with oxygen that can get there pinned at elevated
 temperatures \cite{Senkader2001} and as a function of the annealing time. Materials under 
strain determine a collective motion of dislocations which give rise to strain 
burst difficult to control \cite{Vespignani2001} \cite{Csikor2007}. 
The role of dislocations as a fast route for oxygen diffusion is debated \cite{Metlenko2014}. 
Electro-switching transitions are observed in Mott insulators through the creation of metallic
 paths \cite{Guiot2013}, which might involve oxygen-ion transport along nanoscale dislocation networks. 

Evidence of nanoscale phase separation has recently been found in functional oxides like cobalates \cite{Drees2014}, bismutates \cite{Giraldo-Gallo2015} and in cuprates \cite{Fratini2010, Poccia2011, Campi2015, Carlson2015} . In cuprates, the oxygen interstitials (O-i) organization and the charge density wave (CDW) 
ordering, coupled to the incommensurate periodic local lattice distortions (LLD), have been found to be 
spatially anti-correlated and to influence the macroscopic quantum
 condensate  \cite{Campi2016} \cite{kus} \cite{bia2} \cite{Zaanen2010} \cite{Littlewood2011} \cite{Fratini2010} \cite{Poccia2011} \cite{Poccia2011bis} \cite{Poccia2011tris} \cite{kus2} in a landscape of filamentary networks \cite{Ricci2014} \cite{Poccia2013}  \cite{Campi2015}    \cite{Campi2016} \cite{kus3}.  In these works a non-euclidean geometry has been proposed
  to describe the optimal inhomogeneity of the cuprates \cite{Campi2015} \cite{Campi2016}. Non-euclidean geometries are also known to determine further complexity 
  in the disposition and dynamics of defects \cite{Lipowsky2005}. The nanoscale phase separation  observed in the pnictides suggest
   the occurrence of dislocations at the interface \cite{Ricci2015}. Therefore, gaining control on the network described by the spatial arrangements of ordered defects is important for the design of materials \cite{Aidhy2015}. Indeed, the collective motion of the cooper pairs 
   can be strongly influenced by an annealed complex network \cite{Bianconi2012}. The control 
   of defects is considered a central problem in advanced materials and atomic 
   heterostructure \cite{Boschker2016}, and in the realization of array of artificial atoms 
   in strongly correlated systems \cite{Mannhart2016}. Recently, the 
   precise control of disorder in an array of vortices, which behave as an array 
   of superatoms, has made possible the observation of a dynamic vortex Mott insulator to metal transition \cite{Poccia2015}. 

Here we show that dislocations can act as boundaries between charge density waves and oxygen rich 
puddles in a cuprate superconductor. We use scanning nano X-ray 
diffraction to image the spatial organization of the oxygen interstitials and charge density wave across the dislocation boundary of a cuprate superconductor. 

\section{Experimental method}

The sample of La$_2$CuO$_4$ was grown first by flux method 
and then doped by electrochemical oxidation. The critical superconducting
temperature of the oxygen doped La$_2$CuO$_{4+y}$  ($y=0.12$) 
was determined to be 41 K by single coil resistivity measurements.
The orthorhombic lattice parameters of single crystal were determined to be a=(5.386$\pm$0.004) $\AA$,
b=(5.345 $\pm$ 0.008) $\AA$, c=(13.205 $\pm$0.031) $\AA$ at room temperature. 
The space group of the sample is Fmmm. 
Nano X-ray diffraction experiments were performed on the ID13 beam line of ESRF, 
using a photon energy of 14 keV focused on a 300 x 300 nm$^2$ spot on the sample a-b surface.
X-Ray Diffraction (XRD) patterns were collected in reflection geometry 
to record the reflections on the b*c* reciprocal plane by a Frelon area detector. 
The scanning step was set to 3 $\mu m$ in both horizontal and vertical directions. The experiment has been performed at room temperature.
Thanks to the high brilliance source, it has been possible to record
 a large number of weak superstructural peaks around the main Bragg peaks of the average structure. 

\section{Results and discussion}
Optimally doped La$_2$CuO$_4$$_+$$_y$ is ideal for the
 investigation of interstitials diffusion in copper oxides because the
  O-i are mobile in the La$_2$O$_2$$_+$$_y$ layers 
  intercalated between the superconducting CuO$_2$ planes.

The main Bragg peaks of the single crystal are surrounded by the known satellite peaks with wave-vector q$_O$$_-$$_i$ = 0.09a$^*$ + 0.25b$^*$ + 0.5c$^*$, due to the O-i dopants ordered in three-dimensional superstructure  (see Fig. 1). Alongside the O-i peaks, indexing of superlattice peaks around the Bragg lattice reflections show the presence of incommensurate modulations q$_C$$_D$$_W$ = 0.035a* + 0.21b* + 0.29c* (see Fig. 1). These modulations are charge density waves strongly coupled with the incommensurate periodic local lattice distortions (LLD) of the copper plane \cite{bia2} \cite{Poccia2011} \cite{Campi2015}, forming stripes running in the a direction with a correlation length of about 2.6 nm in the b direction. The correlation between CDW and LLD is know from long time in layered materials \cite{Wilson1975}. These incommensurately modulated LLDs could define the nanoscale organization of CDW domains that show average size of 12 nm. The X ray diffraction profiles of the two satellite reflections, q$_O-$$_i$ and q$_C$$_D$$_W$ in the b*c* plane are shown in Fig. 1. The average increase in the intensity of the q$_C$$_D$$_W$ is observed below 250 K until reaches a maximum at about 100 K \cite{Poccia2013}.

The spatial distribution of both the O-i and the CDW satellites has been studied by using the X-ray micro diffraction apparatus at the ESRF. Scanning the sample using nanometer piezo-stages we recorded the spatial dependence of the O-i and CDW peak intensity. The resulting spatial maps show clear dislocations on the a-b plane of the crystal. Figure 1 shows two typical dislocation running on the a-b plane. Its deepness along the c-axis overcome the X-ray penetration length, while its largeness on the a-b plane extends for less than the beam size (300 x 300) nm.

Our key experimental discovery is that such dislocation can control both the arrangement of the O-i and CDW domains in real space.  It shows that region of CDW higher order and larger clusters are formed on the bottom portion of the image close to the dislocation boundary. On the top part, instead, the O-i order is dominant. More specifically, we observe the enhancement of O-i and the CDW melting in proximity of the dislocation. The intensity profiles along the white six dashed lines of the maps in Fig. 1 are shown in Fig. 2, and highlight the anti-correlated spatial arrangement of the two phases O-i and CDW in the proximity of the dislocation. 

\begin{figure}[h!]
	\begin{center}
		\includegraphics[width=1.0\linewidth]{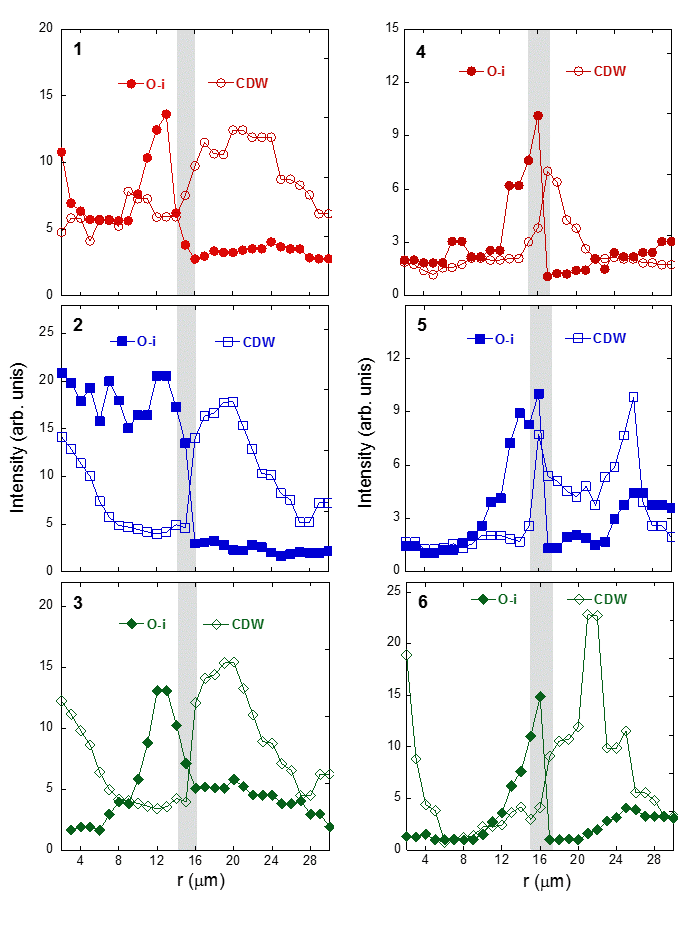}
		\caption{ The profile cut of the integrated intensity map of the q$_O$$_-$$_i$  and q$_C$$_D$$_W$ peak are shown along the dashed white lines as indicated in the maps of Fig. 1. Each panel shows respectively the O-i and CDW intensity along the profile cuts indicated with the numbers from 1 to 6. The grey area is the region occupied by the dislocation.  It is visible the anti-correlation of the O-i and CDW orders in the proximity of the dislocation. 
		}
		\label{symmetry}
	\end{center}
\end{figure}

The general character of this  spatial anti-correlation results evident also from the scatter plot of O-i intensity versus CDW intensity (Fig. 3) calculated on the whole maps of Fig. 1. As O-i intensity becomes higher, the CDW intensity decreases, and vice versa. The observed dislocations determine a physical barrier for the oxygen interstitial diffusivity \cite{Chroneos2011}. The dislocation can be used to manipulate the ions distribution which is considered a relevant problem in modern material science \cite{Kalinin2013}. The concentration of the O-i order on one side the dislocation creates a CDW order on the other side of the dislocation. This artificial separation induced by the dislocation determines a more pronounced spatial anti-correlation between the O-i and the CDW orders.

\begin{figure}[h!]
	\begin{center}
		\includegraphics[width=0.9\linewidth]{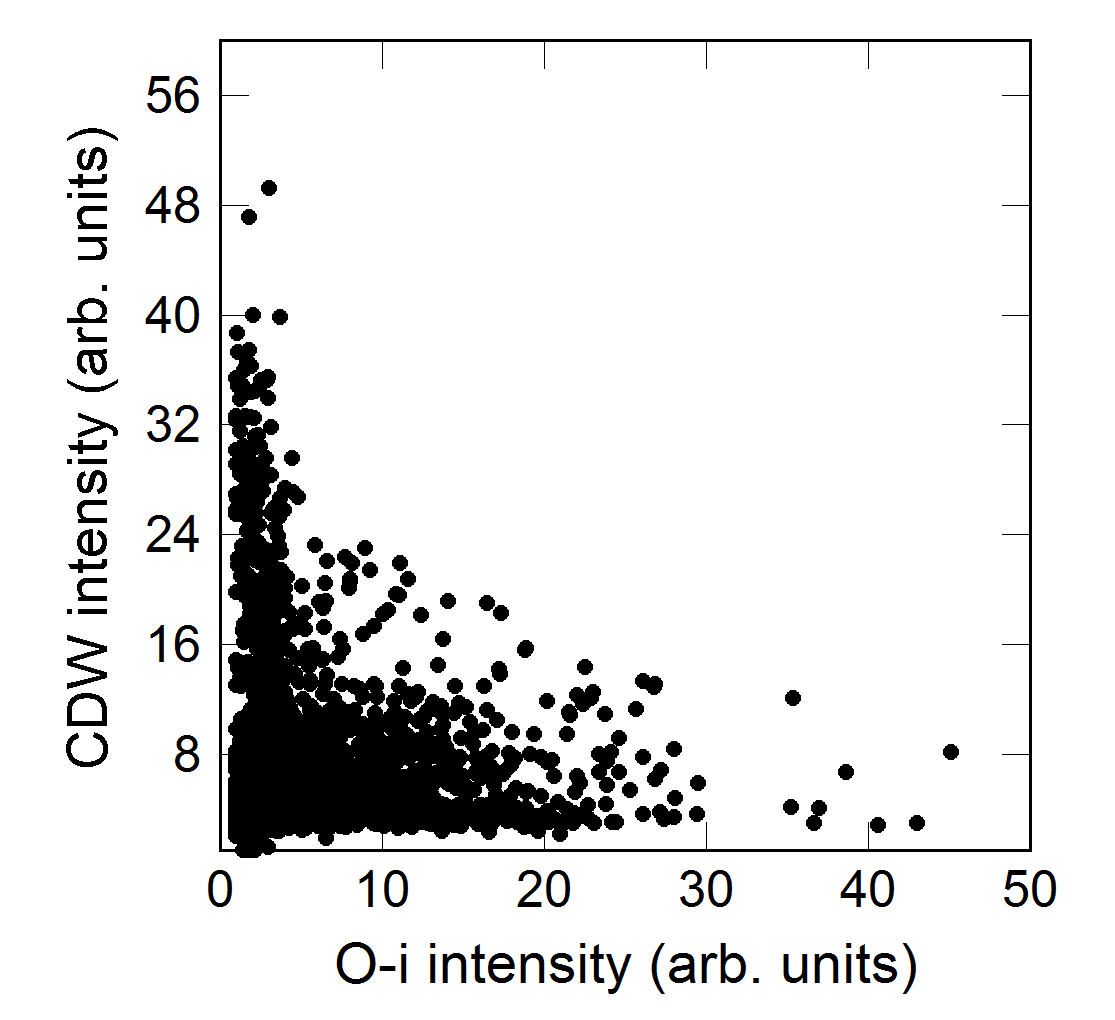}
\caption{The anti-correlation is further quantified by plotting the intensity of the whole maps for q$_C$$_D$$_W$  as a function of q$_O$$_-$$_i$ . The observed scatter plots demonstrate the spatial anti-correlation between the two phases near the boundary.
 	}
 	\label{symmetry}
  \end{center}
\end{figure}

Recently attention on the control and properties of dislocations 
has been given in van der Waals heterostructures \cite{Kushima2015} through 
the observation of atomic scale ripples. Although in the case we report here the dislocation is observed
 in a bulk crystal, the observed pinning of anti-correlated phases could be  basically explained on how  the material accommodate the high load of natural strain.
 Modulation of the local strain in the proximity of the dislocation can determine the observed 
 oxygen ordering \cite{Meng2009}. Elastic field engineering is considered important in many 
field of science, especially focussing on the realization of high - strength 
materials \cite{Zhu2010}. The possibility of modulate the strain field creating dislocations could determine the occurrence of the CDW and O-i rich domains.
It has been suggested that bulk dislocations can modulate the layered solid into ripples and 
that this could be more common in materials than expected \cite{Gruber2016}. 
The modulation of the copper plane as ripples (e.g. short range periodic local lattice distortions) 
close to the dislocation observed in our experiment, can explain the observed pinning 
of the charge density wave order. However, it is also possible to discuss 
the results reported in this paper in terms of a grain boundary separating 
the CDW and the O-i rich domains. It is known indeed that grain boundaries
 can give to the material unique electronic properties different from that of the bulk \cite{Hilgenkamp2002}. 
  
In conclusion, we have performed a scanning nano X-ray 
diffraction experiment and we have imaged the spatial dependence
 of the ordered O-i domains and the CDW rich domains in the proximity of a several tens micrometer 
and diagonal sized dislocations. We observe that the each
dislocation acts as a boundary between the short range incommensurate 
charge density wave in the $CuO_2$ plane, and the oxygen
 interstitials rich domains in the spacer layer corresponding 
 with neighbor filamentary high density metallic wires in the $CuO_2$ plane.
The observed spatial anti-correlation can be exploited in bistable switching
devices thanks to the separation between the two phases. 
The control of dislocations is therefore an approach to control the 
boundaries of phase separation between  O-i rich metallic filamentary domains and CDW rich domain in cuprate oxide superconductors in the so called superstripes  landscape \cite{bia6}.

\section{Acknowledgements}

The work was supported by the Italian Ministry of Education and Research, 
 the Dutch FOM and NWO foundations and Superstripes Institute. 
 We thank ESRF synchrotron facility for support. In particular we thank 
 Manfred Burghammer and the staff of the ID13 beam-line for experimental help.

\end{document}